\documentclass[preprint]{aastex}

\usepackage{rotating}
\usepackage{pdflscape}
\usepackage{mdwlist}

\def \msun   {\hbox{M$_\odot$}}

\shorttitle{Na 8200~\AA~Doublet}
\shortauthors{Schlieder et al.}

\begin{document}

\title{The Na 8200~\AA~Doublet as an Age Indicator in Low-Mass Stars}

\author{Joshua E. Schlieder\altaffilmark{1,2,3}, S\'{e}bastien L\'{e}pine\altaffilmark{4}, Emily Rice\altaffilmark{4}, Michal Simon\altaffilmark{1,2}, Drummond Fielding\altaffilmark{5}, and Rachael Tomasino\altaffilmark{6}}
\altaffiltext{1}{Department of Physics and Astronomy, Stony Brook University,
    Stony Brook, NY 11794, jschlied@ic.sunysb.edu, michal.simon@stonybrook.edu}
\altaffiltext{2}{Visiting Astronomer, NASA Infrared Telescope Facility (IRTF).}
\altaffiltext{3}{Currently at Max-Planck-Institut f\"{u}r Astronomie, K\"{o}nigstuhl 17, 69117 Heidelberg, Germany, schlieder@mpia-hd.mpg.de}
\altaffiltext{4}{Department of Astrophysics,  American Museum of Natural History, Central Park West at 79th Street,
  New York, NY 10024, lepine@amnh.org, erice@amnh.org}
\altaffiltext{5}{Department of Physics and Astronomy,  Johns Hopkins University, 366 Bloomberg Center,
3400 N. Charles Street, Baltimore, MD 21218, dfieldi1@jhu.edu}
\altaffiltext{6}{Department of Physics,  Central Michigan University, Mount Pleasant, MI 48859, tomas1r@cmich.edu}  

\begin{abstract}
 
We investigate the use of the gravity sensitive neutral sodium (\emph{NaI}) doublet at 8183~\AA~and 8195~\AA~(Na 8200~\AA~doublet) as an age indicator for M dwarfs.  We measured the Na doublet equivalent width (\emph{EW}) in giants, old dwarfs, young dwarfs, and candidate members of the $\beta$ Pic moving group using medium resolution spectra.  Our Na 8200~\AA~doublet \emph{EW} analysis shows that the feature is useful as an approximate age indicator in M-type dwarfs with (\emph{V-K$_{s}$}) $\ge$ 5.0, reliably distinguishing stars older and younger than 100 Myr.  A simple derivation of the dependence of the Na \emph{EW} on temperature and gravity supports the observational results. An analysis of the effects of metallicity show that this youth indicator is best used on samples with similar metallicity.  The age estimation technique presented here becomes useful in a mass regime where traditional youth indicators are increasingly less reliable, is applicable to other alkali lines, and will help identify new-low mass members in other young clusters and associations.

\end{abstract}

\section{Introduction}

The members of nearby, young moving groups (\emph{NYMGs}, see Zuckerman \& Song 2004; Torres et al. 2008, hereafter T08) are important because: 1) they represent well characterized samples with close, young stars for the study of their astrophysical properties at $>$3 times the resolution possible in the nearest star forming regions, 2) study of \emph{NYMG} kinematics can infer the recent star formation history in the Solar neighborhood, and 3) gas giant planets are still self-luminous from gravitational collapse at ages 10 - 100 Myr.  This leads to an increased contrast ratio between planet and host.  Thus, \emph{NYMG} members are of great interest to planet hunters (see Lafreni\`ere et al. 2008; Biller et al. 2010; Wahhaj et al. 2011).  However, it is difficult to identify the members of spectral types (\emph{SpTy}) later than M4 because the usual age diagnostics become unreliable.  

West et al. (2008, 2011) showed that the fraction of H$\alpha$ emitting M dwarfs in the Sloan Digital Sky Survey (York et al. 2000) Data Release 7 (Abazajian et al. 2009) spectroscopic sample is strongly dependent on \emph{SpTy}.  Nearly all stars later than $\sim$M4 emit H$\alpha$, while H$\alpha$ emission is only seen in early-type M dwarfs that are young ($\lesssim$1 Gyr).  This is probably a consequence of a change in the magnetic dynamo with the onset of fully convective interiors and also applies to more energetic emission, such as X-rays, since they are often correlated with H$\alpha$ (see Covey et al. 2008).  Li absorption can also serve as an age indicator in low-mass stars, but does not reliably reproduce the expected ages of young cluster members when observations are compared to theory (see Yee \& Jensen 2010).  This discrepancy may be explained by the dependence of Li depletion on stellar accretion history (Barraffe \& Chabrier 2010).  Lithium is also depleted quickly in low-mass stars and may be absent by the ages of the \emph{NYMGs} (Stahler and Palla 2004).      
  
Gravity sensitive photospheric lines, such as those of the alkali metals Na and K, can also indicate a star's age.  It is well known that the neutral sodium (\emph{NaI}) lines in the optical and near infrared are sensitive to photospheric surface gravity in the sense that the stronger the gravity,  the greater the line equivalent width (\emph{EW}).  As the radius of the star decreases on the approach to the main sequence, the photospheric gravity, $g = GM_*/R^2_*$, increases, and the line widths change accordingly.  Alkali atoms are easily pressure broadened because of their single valence electron.  Early studies showed that \emph{NaI} doublet at  $\sim$5900~\AA~(\emph{Na D}) is useful as a luminosity class indicator in late-K and early-M \emph{SpTy's} (Luyten 1923; Spinrad 1962).  Unfortunately TiO absorption renders the \emph{Na D} \emph{EW} useless for these purposes beyond \emph{SpTy's} $\sim$M2.  

Spectroscopic observations of candidate young stars in Taurus, Upper-Scorpius, the Pleiades, and the Solar neighborhood reveal that the \emph{NaI} subordinate doublet at 8183 and 8195~\AA~(Na 8200~\AA) is useful for distinguishing young, mid-M and later-type dwarfs from field dwarfs and giants (Steele \& Jameson 1995, hereafter SJ95; Schiavon et al. 1995; Brice\~no et al. 1998; Mart\'in et al. 2004;  Slesnick et al. 2006a, 2006b, 2008; Takagachi \& Itoh 2010).  These empirical results exhibit sufficient \emph{EW} scatter such that the \emph{NaI} measurement does not provide a precise estimate of age, but does reliably distinguish the stars in young clusters, such as the Pleiades ($\sim$125 Myr), and billion year old stars in the field (Mart\'in et al. 2010, hereafter M10). 
  
Here we concentrate on the doublet at 8200~\AA~(\S 2) and measure the \emph{EW} in the spectra of old dwarfs, giant standards and candidate members of the $\beta$ Pic moving group (\emph{BMPG}, age 10-20 Myr, T08).  We interpret the empirical gravity sensitive behavior of the doublet using calculated, synthetic spectra (\S 3) and line formation of very optically thick, pressure broadened lines in stellar atmospheres (\S 4).  A discussion and summary follow in \S 5 and \S 6.

\section{The Na 8200~\AA~Doublet, Sample Selection, and Measurements}

The \emph{Na D} doublet originates in the $3s$~ground state of \emph{NaI} with wavelengths 5890 \AA~($ 3s^2 S_{1/2} - 3p^2 P_{3/2}$) and 5896 \AA ~($ 3s^2 S_{1/2} - 3p^2P_{1/2}$).  The 8200 \AA ~doublet (\emph{NaI} subordinate doublet) is the transition between  the split $3p$ upper level of the \emph{Na D} lines at 8183 ($3p^2 P_{1/2} - 3d^2D_{3/2}$) and 8195 \AA~($3p^2 P_{1/2} - 3d ^2D_{1/2}$; Fig.~\ref{elevels}; Bashkin \& Stoner 1975).  Fig.~\ref{example_spec} shows a representative sample of low-mass dwarf spectra.  The \emph{Na D} doublet becomes difficult to observe in stars with \emph{SpTy's} later than $\sim$M2 because TiO dominates the spectrum near $\sim$5900~\AA.  The 8200 \AA~doublet is flanked by TiO and VO  absorption bands at shorter and longer wavelengths, and narrow absorption lines of photospheric Fe and telluric H$_2$O  lie just to the red (e.g. SJ95, their Fig. 2).

\subsection{Samples and Observations}

To gauge the gravity sensitivity of the Na 8200~\AA~doublet we used four samples of stars with varying surface gravities.  

\noindent{\emph{High Velocity M Dwarfs}:  Dwarf stars with large transverse velocities (\emph{HV}; V$_{tan}$ $>$ 75 km s$^{-1}$) are kinematically consistent with being members of the thick disk population.  Thus, they are expected to be old ($>$1 Gyr) and have surface gravity log($g$) $\approx$ 5.0 (Drilling \& Landolt 2000, hereafter DL2000).  We chose 203 solar metallicity, \emph{HV} K7 - M9 dwarfs from the LSPM catalogs (L\'epine \& Shara 2005a, hereafter LS05a; L\'epine \& Shara 2005b) for follow up.  The metal content of these stars was determined using the procedures detailed in L\'epine et al. (2007, hereafter L07).  These stars were observed using several different medium resolution spectrographs on different telescopes (S. L\'epine, private communication).
  
\noindent{\emph{Giants}:  The surface gravities of red giants lie in the range 1.0 $\lesssim$ log($g$) $\lesssim$ 3.0 (DL2000).  We selected 4 M giants from the Stony Brook/SMARTS Spectral Standards Library\footnote{http://www.astro.sunysb.edu/fwalter/SMARTS/spstds.html}$^{,}$\footnote{SMARTS, the Small and Medium Aperture Telescope System consortium, under contract with the Associated Universities for Research in Astronomy (AURA)} (Table \ref{giant_tab}).  The giants were observed using the RC Cassegrain Spectrograph on the SMARTS 1.5m telescope at the Cerro Tololo Interamerican Observatory (CTIO) by service observers using grating 58 in the first order.  These settings provide a spectral range 6000 - 9000~\AA~ and a resolution of 6.5~\AA.}

\noindent{\emph{Known $\beta$ Pic Moving Group}:  Members of the 10-20 Myr old \emph{BPMG} are expected to have surface gravities intermediate to giants and old dwarfs.  The radii of these dwarfs are $\lesssim$2 times their main sequence radius.  Therefore, we expect their log($g$)'s to be $\lesssim$0.3 of the field dwarfs.  We chose a sample of 11 late-type members from T08, Schlieder et al. (2010), and Schlieder et al. (2012, in prep.) for follow up (Table \ref{BPMG_tab}).  Observations were made during December 2009 using the the MkIII spectrograph on the 1.3m McGraw-Hill telescope at the MDM Observatory.  We used the 300 line mm$^{-1}$ grating blazed at 8000~\AA~in the first order to produce spectra with a resolution of 6.0~\AA.}  

\noindent{\emph{Young $\beta$ Pic Moving Group Candidates}}:  We studied a subsample of 18 probable young candidates (\emph{PYCs}) selected on the basis of X-ray and UV activity as a youth indicator in Schlieder et al. (2012).  This subsample is comprised of 2 types of \emph{PYCs}, those for which the activity based youth indicators are reliable ((\emph{V-K$_s$}) $<$ 5.0) and those for which activity is an ambiguous youth indicator ((\emph{V-K$_s$}) $\ge$ 5.0).  We refer to the first subsample as candidates with reliable youth indicators (\emph{CWRYs}) and the latter as candidates with ambiguous youth indicators (\emph{CWAYs}).  These candidates are listed in Table \ref{PYC_tab} and were observed with the known members in 2009 December.}   


\subsection{Na 8200~\AA~\emph{EW} Measurements}

All of the MDM spectra were reduced using IRAF.  A NeAr arc lamp spectrum was used to calibrate the wavelength scale and flux calibration was determined from observations of spectrophotometric standards from Oke (1990).  \emph{SpTy}'s for all dwarf samples were determined using the methods described in L\'epine et al. (2003).  The giant standard spectra were reduced by F. Walter using his pipeline procedures (Walter et al. 2004).  Giant \emph{SpTy}'s were taken from the Stony Brook/SMARTS Spectral Standards Library.  The telluric H$_2$O feature that contaminates the 8200~\AA~doublet was corrected in all spectra using custom software written in IDL (S. L\'epine, priv. comm.). If uncorrected, this feature can increase the measured \emph{EW} of the doublet by $\sim$10\% (SJ95)    

The Na 8200~\AA~doublet becomes prominent around \emph{SpTy} M0 (see Fig.~\ref{example_spec}).  We placed all spectra on the same 3~\AA~resolution and measured the doublet in all stars where it was apparent in a visual inspection of the spectra.  Since the Na 8200~\AA~doublet is unresolved in our spectra, we chose a central wavelength ($\lambda$$_c$) of 8190~\AA~and integrated the flux in a 22~\AA~region around $\lambda$$_c$.  To normalize the line width, we calculated the average flux in pseudo-continuum regions blueward and redward of the doublet. For all dwarf samples these regions were 8149~\AA~- 8169~\AA~and 8236~\AA~- 8258~\AA.  Slightly different continuum regions were chosen for the giants but $\lambda$$_c$ remained the same.   We maintained the same measurement procedure, identical $\lambda$$_c$'s, and nearly identical continuum regions despite the diverse range of instruments used.  This minimizes the systematic uncertainty introduced by the use of different instruments. Our code allowed widths to be measured on a star by star basis while monitoring signal to noise, thereby avoiding particularly noisy measurements.


Fig.~\ref{82_SDF} shows the Na 8200~\AA~\emph{EW} as a function of (\emph{V-K$_s$}) for the observed samples.  The \emph{V} magnitudes of the dwarf samples were calculated from USNO1.0-B (Monet et al. 2004) \emph{B}, \emph{R}, and \emph{I} values following the methods described in LS05a, \emph{K$_s$} magnitudes were drawn from 2MASS (Skrutskie et al. 2006).  Fig.~\ref{82_SDF} shows a second order polynomial fit to the \emph{HV} M dwarf data (dashed gray line) with measurement uncertainties.  The individual data for the \emph{HV} M dwarfs is shown in Fig.~\ref{metal_plot} and discussed in \S 5.  The isochrones shown in Fig.~\ref{82_SDF} are described in \S 4.  For stars having (\emph{V-K$_s$}) $\gtrsim$ 5.0 we find a clear distinction between the Na 8200~\AA~line \emph{EW} of young and old samples.  A Kolmogorov-Smirnov ({\emph{KS}) test yields a negligibly small probability that the two samples (the young dwarfs shown in Fig.~\ref{82_SDF} and the old M-dwarfs with solar metallicity shown in Fig.~\ref{metal_plot}) were drawn from a common population and differ only by the noise of sampling, of order 10$^{-3}$.  The 2 known BPMG members with (\emph{V-K$_s$}) $>$ 5.0 and several of our \emph{CWAYs} have \emph{EW}s which are less than those of the old \emph{HV} dwarfs (see Tables \ref{PYC_tab}).  Conversely, a few \emph{CWAYs} have \emph{EW}s inconsistent with known young stars.  This indicates that the Na 8200~\AA~\emph{EW} is useful in differentiating old, active stars from those that are truly young.  Johnson and Apps (2009) suggest metallicity has an effect on both the \emph{EW} and the (\emph{V-K$_s$}) color.  We discuss the effects in the following section and in \S 5.   

\section{Comparison to Models}

We compared our observations to model predictions using Siess et al. (2000, hereafter SDF2000)\footnote{See http://www-astro.ulb.ac.be/$\sim$siess/index.html} and Baraffe et al. (1998, hereafter BCAH98)\footnote{See http://perso.ens-lyon.fr/isabelle.baraffe/} solar metallicity, pre-main sequence (\emph{PMS}) evolutionary models combined with synthetic spectra generated using the PHOENIX stellar atmospheres code (Hauschildt et al 1999; Rice et al. 2010; hereafter R10).  We obtained effective temperature (\emph{T$_{eff}$}) and log($g$) data for grids of \emph{PMS} models with ages 1 Myr - 1 Gyr and masses 0.1 M$_{\odot}$ - 0.5 M$_{\odot}$.  To measure the \emph{EW} of the 8200~\AA~doublet for the models we produced theoretical spectra for our (\emph{T$_{eff}$}, log($g$)) grid using PHOENIX.  We reduced the spectra to 3~\AA~resolution and measured the doublet \emph{EW} in the same way as the observed dwarf spectra.  Fig. \ref{syn_spec} shows \emph{T$_{eff}$}=3300K theoretical spectra with 4 log($g$)'s which approximately represents the contraction of an 0.2 M$_{\odot}$ star to the main-sequence.  The increase in Na line width with log($g$) is apparent. 

In Fig.~\ref{obs_comp} we show trimmed, convolved PHOENIX synthetic spectra compared to observations across the entire observed wavelength range. In the top of the figure, observed spectra (black) of old, weakly active dwarfs are compared to 1 Gyr old model spectra (red).  A likely young, mid-M dwarf is compared to a 10 Myr old model (green) in the bottom of the figure.  The chosen observed spectra have temperatures consistent with the models using the \emph{T$_{eff}$} to \emph{SpTy} conversion of Kenyon and Hartmann (1995, hereafter KH95)\footnote{See also the conversion table by Eric Mamajek at http://www.pas.rochester.edu/$\sim$emamajek} for the main sequence stars and the conversion scale of Luhman et al. (2003) for the young dwarf (see Table \ref{SDFmod}).  The figure shows that for both main sequence and young M dwarfs the observed and model spectra match both the continuum and absorption features well across the observed wavelength range.  There are deviations in some of the molecular bands due to missing opacity in the models, particularly the TiO band at $\sim$5900~\AA.  This arises from the incomplete treatment of complex molecular line lists suffered by the models (Chabrier and Baraffe 1997; BCAH98).  The discrepancy in the TiO band increases as mass decreases in the main sequence dwarfs.  However, the match is better at this feature in the young M5; this is consistent with the results of R10 for high resolution J-band spectra.  In their analysis of PHOENIX model fits to late-M dwarf spectra in the near-IR they find the molecular features are better represented in young, low-gravity models.  The comparison in the region around the Na 8200~\AA~doublet is consistent.  We compared the measured \emph{EWs} of several observed spectra and their synthetic counterparts.  The \emph{EW} differences for the spectra shown in Fig.~\ref{obs_comp} are +0.4, +0.3, -1.0 and 0.0~\AA.  For several other comparisons (not shown) the differences were in the range $\pm$0.2~\AA.  The scatter is less than
that among the measured \emph{EWs} as shown in Figs.~\ref{82_SDF} and \ref{metal_plot}, and less than the differences attributable to gravity at (\emph{V-K$_s$}) $\gtrsim$ 5.0.  This simple visual comparison shows that the observed spectra and models match well not only in the region of interest around the Na 8200~\AA~doublet, but across red, optical wavelengths.  Our comparison is motivation for a full analysis of optical model fits to observed low-mass dwarfs spanning a range of masses, ages, activity levels, and metallicities.  This analysis is the focus of another paper (Schlieder et al., in prep.).        

The SDF2000 models provide \emph{V} and \emph{K} magnitudes from the empirical \emph{T$_{eff}$} to color conversion of KH95 (see Table \ref{SDFmod}).  We checked the conversion from the model \emph{K} mags to \emph{K$_{s}$} mags\footnote{KH95 use Bessel and Brett (1988) near-IR colors for K - M dwarfs, we used the conversion from \emph{K$_{BB}$} to  \emph{K$_{s}$} given at http://www.astro.caltech.edu/$\sim$jmc/2mass/v3/transformations/.  See also the magnitude conversion by Eric Mamajek at http://www.pas.rochester.edu/$\sim$emamajek/memo\_BB88\_Johnson2MASS\_VK\_K.txt} and found the average difference to be $\sim$0.04 mags, which is insignificant in our analysis.  In Fig.~\ref{82_SDF} we show only SDF2000 isochrones because the BCAH98 models are consistent when their \emph{T$_{eff}$}'s are used to interpolate magnitudes from the KH95 scale.  We do not show the 1 Myr, 100 Myr, and 1 Gyr data point for the 0.1 M$_{\odot}$ model because the predicted temperatures at these ages are beyond the limits of the KH95 conversion scale.  


A star's metallicity affects both its 8200~\AA~doublet \emph{EW} and (\emph{V-K$_s$}) color.  The Na 8200~\AA~doublet is saturated, hence its \emph{EW}$\propto$(\emph{NaI})$^{1/2}$, and to a first approximation, \emph{Z}$^{1/2}$, where \emph{NaI} is the number density of neutral sodium, and \emph{Z} the metallicity.  We used SDF2000 models to obtain colors at \emph{Z} = 0.01, 0.02, and 0.04 for 0.5 and 0.2 M$_{\odot}$ stars at 1 Gyr age.  The shifts in the ((\emph{V-K$_s$}), \emph{EW}) plane are indicated by the orange vectors in Fig.~\ref{82_SDF}; the dots indicate the extremes of metallicity, twice solar and half solar.  The increment in \emph{EW} is given by its (\emph{NaI})$^{1/2}$ dependence.  The increment in (\emph{V-K$_s$}) is determined by the change in \emph{T$_{eff}$} with metallicity in SDF2000 models.  Even small differences in metallicity can produce substantial shifts in the ((\emph{V-K$_s$}), \emph{EW}) plane; we discuss the effects of metallicity further in \S 5.

\section{Temperature and Gravity Dependence of Na 8200~\AA}

In this section we address the following questions:  1) Why does the Na 8200~\AA~doublet become prominent in stars later than M0 \emph{SpTy}?  2) How does the strength of the doublet depend on photospheric gravity?  3) Why does the \emph{EW} of the doublet become useful as a youth indicator for (\emph{V-K$_s$}) $\gtrsim$ 5.0?

We used the PHOENIX code to calculate line profiles and evaluate the temperature and gravity dependence of the Na doublets in low-mass stars.  To calculate the spectra, PHOENIX  derives the structures of the atmospheres in terms of temperature (\emph{T}), gas pressure (\emph{P$_g$}), electron pressure (\emph{P$_e$}), and optical depth (\emph{$\tau$}).  We calculated two sets of models:  1) Stars of mass 0.7, 0.5, 0.4, 0.2, and 0.1 \msun~at 1 Gyr age, 2) stars of mass 0.7, 0.2, and 0.1 \msun~at ages 1, 10, 50, and 100 Myr, and thus model their photospheres as they contract.


The fraction of Na that remains neutral in the atmospheres of these stars is given by the Saha equation    

\begin{equation}
{{Na II}\over {Na I}} N_{e} = 2 {{Z_{II} \over {Z_{I}}}} \left({{2\pi m_{e}kT}\over {h^2}}\right)^{3 \over 2 } e^{- {\xi_i \over {kT}}}, 
\end{equation}

\parindent=0.0in
where $Na II, Na I,$ and $N_{e}$ are the densities of singly ionized and neutral sodium, and  electrons.  With first ionization potential $\xi_i = 5.1$ eV,  $P_{e} = N_{e} kT$ and the ratio of partition functions  $Z_{II} /Z_I = 1.0/2.0$ (Cayrel \& Jugaku 1963),

\begin{equation}
{{Na II}\over {Na I}}   = 0.33 {T^{5\over 2} \over P_{e}} e^{-{59600\over T}}.
\end{equation}

The second ionization potential of Na is 47 eV; hence at the temperatures of these stars, $Na III/Na II$ is negligibly small and, to an excellent approximation, the total abundance of Na is $Na_{tot} = Na I + Na II $.  Fig. \ref{abunrat} shows the results as a function of continuum optical depth in 1 Gyr old stars;  as $T_{eff}$ of the star decreases an increasingly larger fraction of Na remains neutral.  This explains the increasing strength of the 8200~\AA~doublet in stars cooler than M0.   The increasing fraction of neutral Na compensates for the precipitous decrease of the population in the 3p to 3s levels.  The Boltzmann equation, 

\begin{equation}
{NaI(3p)\over { NaI(3s)}}=  {g_{3p}\over g_{3s}} e^{-{\Delta E\over {kT}}}\propto e^{-{24400\over T}},
\end{equation}

shows that,

$$ {NaI(3p)\over { NaI(3s)}}= {1 \over {202}} ~~to~~ { 1\over {1500}}, $$ 

for \emph{T} = 4000 to 3000  K.  The \emph{Na D} lines would also appear very strong at spectral types cooler than M0 were it not for their obscuration by TiO, since essentially all \emph{NaI} is in the ground state,

\begin{equation}
NaI(3s)  \approx  {{NaI} \over {Na_{tot}}} Na_{tot}   \propto P_{gas}.
\end{equation}

\parindent=0.5in
To calculate the strength of the 8200 \AA~doublet in absorption we follow the approach of Gray (1992, hereafter G92) and write for the increment of opacity in the line, $\tau_{line}$, 

\begin{equation}
d \tau_{line} = (\kappa_{cont} + \kappa_{line}) dx = \left( 1 + {\kappa_{line} \over \kappa_{cont}} \right) \kappa_{cont} dx,
\end{equation}
\parindent=0.0in

where $\kappa_{line}$ is the absorption coefficient in the 8200 \AA~doublet and $\kappa_{cont}$ is the absorption coefficient in the continuum near the lines.  The opacity in the line plus continuum is necessarily greater than in the continuum alone and the line forms higher in the stellar photosphere than the continuum.  Since the temperature is lower at the higher levels the line appears in absorption.  In low-mass stars such as these, $\kappa_{cont}$ is bound-free absorption of H$^-$.  It dominates except for wavelength regions where absorption bands such as those of  TiO and VO are strong.   Absorption by H$_2$O becomes important below $T_{eff} \sim$ 2500K (Auman 1969).   The bound-free absorption coefficient of $H^-$ evaluated at $\lambda=8200$ \AA~is (G92, eqns 8.11 and 8.12)

\begin{equation}
\kappa_{cont}(H^-, 8200~\AA) = 0.194 {P_{e}\over T^{5/2}} 10^{3800/T}{P_{gas}\over T} \propto  P_{e} P_{gas} h(T),
\end{equation}

where $h(T)$~contains the temperature dependence.

\parindent=0.5in

The line absorption coefficient is (G92, eqn 11.12 et seq)

\begin{equation}
\kappa_{line} = {{\pi e^2} \over {m_{e} c^2}} N_l f_{l,u}  \phi (\nu-\nu_o), 
\end{equation}
\parindent=0.0in

where $N_l$ is the population of the lower level, here $N(3p)$, $f_{l,u}$ is the lower to upper level oscillator strength, and $\phi$ is the line shape function, normalized such that $\int \phi (\Delta \nu) d( \Delta \nu )= 1$, where $\Delta \nu = \nu - \nu_0$ and $\nu_0$ is the frequency of the center line.

\parindent=0.5in

The \emph{Na D} and 8200~\AA~doublets are optically  thick (see Fig. \ref{syn_spec}).  The very large line widths, several decades larger than the thermal widths, are produced by van der Waals broadening (Curtis \& Jefferies 1968; Burrows \& Volobuyev 2003).  It is conventional to designate the van der Waals ``damping parameter'' as  $\gamma_6$ (e.g. G92, Eqn. 11.34 et seq.).  Far in the wings  of the line the first term in the Voigt profile suffices,

\begin{equation}
\phi_6 = {0.564\over \pi^{1\over2}} \gamma_6 \left({1\over {\Delta\nu}}\right)^2.
\end{equation}
\parindent=0.0in

The van der Waals damping term is

\begin{equation}
\gamma_6 = 1.6 \times 10^7 {P_{gas} \over T^{7\over {10}}}
\end{equation}

(Uns\"old 1955).  
 

We combine Eqns. (3-4),(6-8), and (9) and find 

\begin{equation}
\kappa_{line} \propto P_{gas}^2  f(T),
\end{equation}

where $f(T)$~represents the temperature dependence.   Hence,

\begin{equation}
{\kappa_{line}\over \kappa_{cont}}  \propto   {P_{gas}\over P_{e}}{{{f(T)}}\over {{h(T)}}}.
\end{equation}
\parindent=0.5in

We used the BCAH98 models for 0.1,  0.2, and 0.7 \msun~stars at 1, 10, 50, and 100 Myr age to determine the dependence of $P_{gas}/P_{e}$~ on photospheric gravity $g$~ at $\tau_{cont}(1.2 \mu m) =1$.  Fig. \ref{gasrat} shows that $P_{gas}/P_{e}$ increases with $g$ in the photosphere and hence with increasing age and explains the gravity and temperature dependent nature of the doublet strength.  Further, the dependence of $P_{gas}/P_{e}$ on $g$ strengthens as mass decreases, a result consistent with the empirical demonstration in Fig.~\ref{82_SDF}.  


\section{Discussion}

\subsection{Effect of Metallicity on \emph{EW} and Color}



Fig.~\ref{metal_plot} shows the Na 8200~\AA~\emph{EW} as a function of (\emph{V-K$_s$}) for 4 samples of M dwarfs with increasingly smaller metallicity.  In the figure, the solar metallicity, \emph{HV} M dwarfs are represented by black dots.  The second order polynomial fit to their distribution (gray, dashed line) is described in the caption.  The three other samples are K7 - M8 sequences of low metallicity, subdwarf, extreme subdwarf, and ultrasubdwarf standards from L07.  L07 determined their metallicities from the observed TiO/CaH ratio.  The ultrasubdwarfs represent the most metal poor M dwarfs in the galactic halo.


The change in \emph{EW} and color is apparent, lower metallicity leads to smaller Na \emph{EWs} and bluer (\emph{V-K$_s$}) colors, consistent with the predictions in \S 3.  The extreme and ultrasubdwarf samples occupy the same area in the Fig.~\ref{metal_plot} and can be regarded as one sample within measurement errors.  The \emph{KS} test applied to the M dwarfs and subdwarfs indicates a high probability, \emph{P}=0.64, that they are drawn from the same population.  The test also shows that the extreme and ultrasubdwarfs have negligible probabilities, \emph{P} $\sim$10$^{-6}$ and 10$^{-2}$, of being drawn from the same sample as the M dwarfs and subdwarfs respectively.  Thus, the application of the Na 8200~\AA~\emph{EW} to estimate age is only reliable for samples of homogeneous metallicity.


Of the stars observed for the Na 8200~\AA~age indicator analysis, only the \emph{HV} M dwarfs have measured metallicity.  However,  all of the $\beta$ Pic moving group members and candidates have kinematics consistent with the solar neighborhood and most are known to be young.  Thus, we expect their metallicities to be approximately solar or super-solar.  If these dwarfs are very metal rich, based on predictions, we expect their positions in Fig.~\ref{82_SDF} to be systematically shifted to the upper right.  This is not consistent with the observations.  We therefore conclude that the dwarfs used in the age indicator analysis represent a homogeneous, approximately solar metallicity sample in which stars with systematically small Na 8200~\AA~\emph{EWs} are young.



\subsection{Application of Results to Other Alkali Lines}

The gravity and temperature sensitivity of the optical Na doublets discussed here is applicable to the other \emph{NaI} lines in Fig.~\ref{elevels} and other alkali atoms and ions with similar electronic structure.  For example, Greene and Lada (1996) show that the combined \emph{EW}s of \emph{NaI} at $\sim$2.2 $\mu$m and \emph{CaI}  triplet around 2.3 $\mu$m in young stellar objects in $\rho$ Oph are intermediate to main sequence dwarfs and giants, consistent with our results for the Na 8200~\AA~doublet.  In addition,  McGovern et al. (2004) show that \emph{KI}, \emph{RbI}, and \emph{CsI} in the optical and \emph{KI} in the near-IR J-band exhibit the same gravity sensitivity discussed here.  These features are useful as youth diagnostics in the very cool photospheres of brown dwarfs.  


\section{Summary}

We present an analysis of the Na 8200~\AA~doublet in several stellar samples to show its usefulness as an age indicator for dwarfs of mid-M \emph{SpTy} and later.  The samples were chosen to be representative of different stages of stellar evolution, from the \emph{PMS} to evolved giants, which in turn trace the evolution of photospheric gravity.  We measured the \emph{EW}s of the 8200~\AA~doublet and compared the empirical results to isochrones generated using BCAH98 and SDF2000 evolution models and the PHOENIX model atmosphere code.  We find that the Na 8200~\AA~doublet is useful as an approximate discriminant of youth for stars with (\emph{V-K$_s$}) $\gtrsim$ 5.0, or $\sim$M4 \emph{SpTy} and later.   The empirical results are consistent with previous studies and with the theoretically derived dependence of the doublet \emph{EW} on temperature and gravity.  The \emph{EW} of the doublet is useful for mid-M and later dwarfs because:  1) it becomes prominent in the coolest dwarfs because of the increased abundance of neutral Na and 2) the dependence on gravity is stronger in 0.1 M$_{\odot}$ dwarfs than in 0.7 M$_{\odot}$ dwarfs.  Analysis of the effects of metallicity show that this youth indicator is best used on samples with similar metallicity.  The \emph{EW} technique as we have described it cannot measure an age.  It can nonetheless discriminate between old M dwarfs in the field and relatively young ones.  As such, we can expect the technique can be profitably applied to identify new, low-mass members of young clusters and associations.

\
\
\
\acknowledgments

We thank the referee for an exceptionally helpful and prompt report.  We thank T. Barman for many useful discussions.  J.E.S. thanks T. Herbst for helpful discussions.  The work of J.E.S and M.S. was supported in part by NSF grant AST 09-07745.  The work of S.L. was supported by NSF grants AST 06-07757 and AST 09-08406.  The work of D.F. and R.T. was supported by NSF REU site grant PHY-0851594.  This publication makes use of data products from the Two Micron All Sky Survey, which is a joint project of the University of Massachusetts and the Infrared Processing and Analysis Center/California Institute of Technology, funded by the National Aeronautics and Space Administration and the National Science Foundation.  This research has made use of the SIMBAD database, Aladin, and Vizier, operated at CDS, Strasbourg, France.

\clearpage

\begin{table}
\begin{center}
\caption{\textbf{\label{giant_tab}Giant Standards\newline}}
\begin{tabular}{lcccccc}
\hline
\hline
Name		       &$\alpha$$^{a}$		          &$\delta$$^{a}$		&\emph{V}	           &\emph{K$_s$}                 &\emph{SpTy}          &EW\\			       
			       &(2000.0)			&(2000.0)		&(mag)        &(mag)           &                    &(\AA)\\
\hline
\hline
HD 1879		      &5.768131	  		&-15.942617  	&6.5        &2.3               &M2III               &1.8\\
HD 27598               &65.172214 		&-16.830004  	&7.1        &1.8               &M4III               &1.8\\
HD 198026             &311.934330 	         &-5.0276032  	&4.5        &-0.3               &M3III               &1.3\\
HD 207076             &326.632640  	         &-2.212806  	&6.8        &-1.7               &M7III               &0.6\\
\hline											
\multicolumn{7}{l}{\scriptsize{$^{a}${ICRS epoch J2000.0 RA and Dec in decimal degrees}}}\\
\end{tabular}
\end{center}
\end{table}		

\clearpage

\begin{table}
\begin{center}
\caption{\textbf{\label{BPMG_tab}\emph{BPMG} Members\newline}}
\begin{tabular}{lcccccc}
\hline
\hline
Name		       &$\alpha$$^{a}$			          &$\delta$$^{a}$		&\emph{V}	           &\emph{K$_s$}                 &\emph{SpTy}          &\emph{EW}\\			       
			       &(2000.0)			&(2000.0)		&(mag)        &(mag)           &                    &(\AA)\\
\hline
\hline
TYC 1208 468 1$^b$      &24.414167  		&18.592500  	&10.7        &6.7                 &K7V                &1.6\\
HIP 11152$^b$                &35.860833  	         &22.735278  	&11.6        &7.3               &M1V               &2.1\\
AG Tri$^d$                        &36.872083  	         &30.973611  	&10.2        &7.1               &K7V               &1.7\\
BD+05 378$^d$               &40.357917  	         &5.988333  	&10.2       &7.1               &K7V                &1.7\\
PM I04439+3723$^b$    &70.987083  	         &37.384250  	&13.4        &8.8               &M2V               &3.1\\
V 1005 Ori$^d$                &74.895000  	         &1.783611  	&10.3        &6.3               &M0V               &2.0\\
TYC 1281 1672 1$^b$   &75.205375 		&15.450194  	&11.0        &7.6               &K7V                &1.8\\
PYC J05019+0108$^c$    &75.485625  	         &1.145250  	&13.2        &7.7               &M5V               &3.3\\
V 1311 Ori$^d$                &83.018750  	         &-3.091389  	&11.4        &7.0               &M2V               &2.7\\
PM I07295+3556$^b$    &112.379500  	         &35.933389  	&12.1        &7.8               &M1V               &2.5\\
PYC J21376+0137$^c$    &324.417500  	         &1.620556  	&13.6        &7.9               &M5V               &3.6\\
BD-13  6424$^d$            &353.128750  	         &-12.264444  	&10.9        &6.6               &M0V               &2.2\\
\hline											
\multicolumn{7}{l}{\scriptsize{$^{a}${ICRS epoch J2000.0 RA and Dec in decimal degrees}}}\\[-0.08 in]
\multicolumn{7}{l}{\scriptsize{$^{b}${Schlieder et al. (2010)}}}\\[-0.08 in]
\multicolumn{7}{l}{\scriptsize{$^{c}${Schlieder et al. (2012, in prep.) }}}\\[-0.08 in]
\multicolumn{7}{l}{\scriptsize{$^{d}${Torres et al. (2008)}}}\\
\label{BPMG_tab}
\end{tabular}
\end{center}
\end{table}

\begin{table}
\begin{center}
\caption{\textbf{\label{PYC_tab}Probable Young Candidates$^{a}$\newline}}
\begin{tabular}{lcccccc}
\hline
\hline
\multicolumn{7}{c}{\textbf{\emph{CWRYs}}}\\
\hline 
\hline 
Name		       &$\alpha$$^{b}$		          &$\delta$$^{b}$		&\emph{V}	           &\emph{K$_s$}                 &\emph{SpTy}          &\emph{EW}\\			       &(2000.0)			&(2000.0)		&(mag)        &(mag)           &                    &(\AA)\\
\hline
PYC J03108+1838    &47.723206  		&18.644018  	&12.4	&8.8   	     &M0		&1.8\\
TYC 3339 1107 2   &59.333015  		&50.855167  	&10.6        &7.3               &K7               &1.3\\
PYC J03575+2445    &59.391407        	&24.752939  	&13.0   	&8.7   	     &M2		&2.3\\
HIP 47133	       &144.066330		&37.529320	&11.1         &7.2              &M2               &2.4\\
PYC J10571+0544   &164.297470               &5.748539        &13.2         &9.0              &M2                &2.6\\
PYC J11519+0731   &177.986725               &7.523960        &12.7         &7.9              &M3                &4.3\\
\hline\hline
\multicolumn{7}{c}{\textbf{\emph{CWAYs}}}\\
\hline 
\hline 							              
PYC J02179+1225$^{c}$     	&34.483406  	&12.424052  	&14.4     	&9.1    		&M4			&2.6\\
PYC J03047+2203$^{c}$     	&46.183666  	&22.055891  	&15.9     	&9.7    		&M5			&3.4\\
PYC J04116+2504$^{c}$  	&62.901592  	&25.078272  	&14.8  	&9.4   		&M4			&2.5\\
PYC J04147+3816$^{c}$  	&63.691860  	&38.267742  	&15.7  	&9.8     		&M5			&3.3\\
PYC J07266+1850  	&111.673080  	&18.842928  	&14.2  	&9.1     		&M4			&3.5\\
PYC J08227+0757  	&125.697861   	&7.954772  	&14.6   	&9.2      		&M4			&4.1\\
\hline
\multicolumn{7}{l}{\scriptsize{$^{a}${Based on strong X-ray/UV activity and (\emph{V-K$_s$}) color, see Schlieder et al. (2012)}}}\\[-0.08 in]
\multicolumn{7}{l}{\scriptsize{$^{b}${ICRS epoch J2000.0 RA and Dec in decimal degrees}}}\\[-0.08 in]
\multicolumn{7}{l}{\scriptsize{$^{c}${Candidates consistent with youth in Fig.~\ref{82_SDF}}}}\\
\label{PYC_tab}
\end{tabular}
\end{center}
\end{table}		

\clearpage

\clearpage

\begin{table}
\begin{center}
\begin{scriptsize}
\caption{\textbf{\label{SDFmod}SDF2000 \emph{Z} = 0.02 Model Data$^a$\newline}}
\begin{tabular}{lcccc}
\hline
\hline
\multicolumn{5}{c}{\textbf{Pre-Main Sequence Stars}}\\
\hline  
\hline 
\multicolumn{5}{c}{0.5~\msun}\\
\hline  
Age		& \emph{T$_{eff}$}			& log$(g)$	&{EW}		&(\emph{V-K})$^{b}$\\		              
(Myr)		&  (K)					&			&(\AA)		&(mag)\\			 	  
\hline	
1		&3771				&3.5			&1.6		&3.8\\
5		&3760				&4.0			&1.7		&3.8\\
10		&3764				&4.2			&1.7		&3.8\\
50		&3796				&4.5			&1.8		&3.7\\
100		&3828				&4.7			&1.9		&3.6\\
\hline
\multicolumn{5}{c}{0.3~\msun}\\
\hline  
Age		& \emph{T$_{eff}$}			& log$(g)$	&{EW}		&(\emph{V-K})$^{b}$\\
(Myr)		&  (K)					&			&(\AA)		&(mag)\\			 	  
\hline	
1		&3360				&3.5			&1.9		&5.3\\
5		&3429				&4.0			&1.9		&4.9\\
10		&3475				&4.2			&2.0		&4.7\\	
50		&3531				&4.5			&2.2		&4.4\\
100		&3529				&4.7			&2.5		&4.4\\
\hline
\multicolumn{5}{c}{0.2~\msun}\\
\hline
Age		& \emph{T$_{eff}$}			& log$(g)$	&{EW}		&(\emph{V-K})$^{b}$\\			              
(Myr)		&  (K)					&			&(\AA)		&(mag)\\			 	  	
\hline
1		&3075				&3.5			&2.3		&7.1\\
5		&3235				&4.0			&2.3		&6.2\\
10		&3290				&4.2			&2.4		&5.8\\	
50		&3330				&4.7			&2.9		&5.5\\
100		&3317				&4.9			&3.6		&5.6\\
\hline
\multicolumn{5}{c}{0.1~\msun}\\
\hline
Age		& \emph{T$_{eff}$}			& log$(g)$	&{EW}		&(\emph{V-K})$^{b}$\\			              
(Myr)		&  (K)					&			&(\AA)		&(mag)\\				 	  	
\hline
1		&2928				&3.5			&$\cdots$		&$\cdots$\\
5		&3023				&4.0			&3.0		&7.5\\
10		&3075				&4.2			&3.3		&7.1\\	
50		&3055				&4.7			&5.0		&7.3\\
100		&3000				&4.9			&$\cdots$		&$\cdots$\\
\hline
\hline
\multicolumn{5}{c}{\textbf{1 Gyr Stars}}\\
\hline
\hline
\emph{SpTy}$^b$		& \emph{T$_{eff}$}			& log$(g)$	&EW		&(\emph{V-K})$^{b}$\\			              	
                           &  (K)					&			&(\AA)		&(mag)\\	
\hline		 	  	
M0			&3804				&4.9		&1.9		&3.7\\
M2			&3669				&5.0			&2.8		&4.5\\
M4			&3290				&5.1			&4.2		&5.6\\
M6			&2780				&5.3			&$\cdots$		&$\cdots$\\
\hline
\multicolumn{5}{l}{\tiny{$^{a}${From Siess et al. (2000) models}}}\\
\multicolumn{5}{l}{\tiny{$^{b}${From KH95, \emph{K} is consistent with \emph{K$_s$}}}}\\
\label{SDFmod}
\end{tabular}
\end{scriptsize}
\end{center}
\end{table}

\clearpage

\clearpage

\begin{figure}
\epsscale{1.0}
\plotone{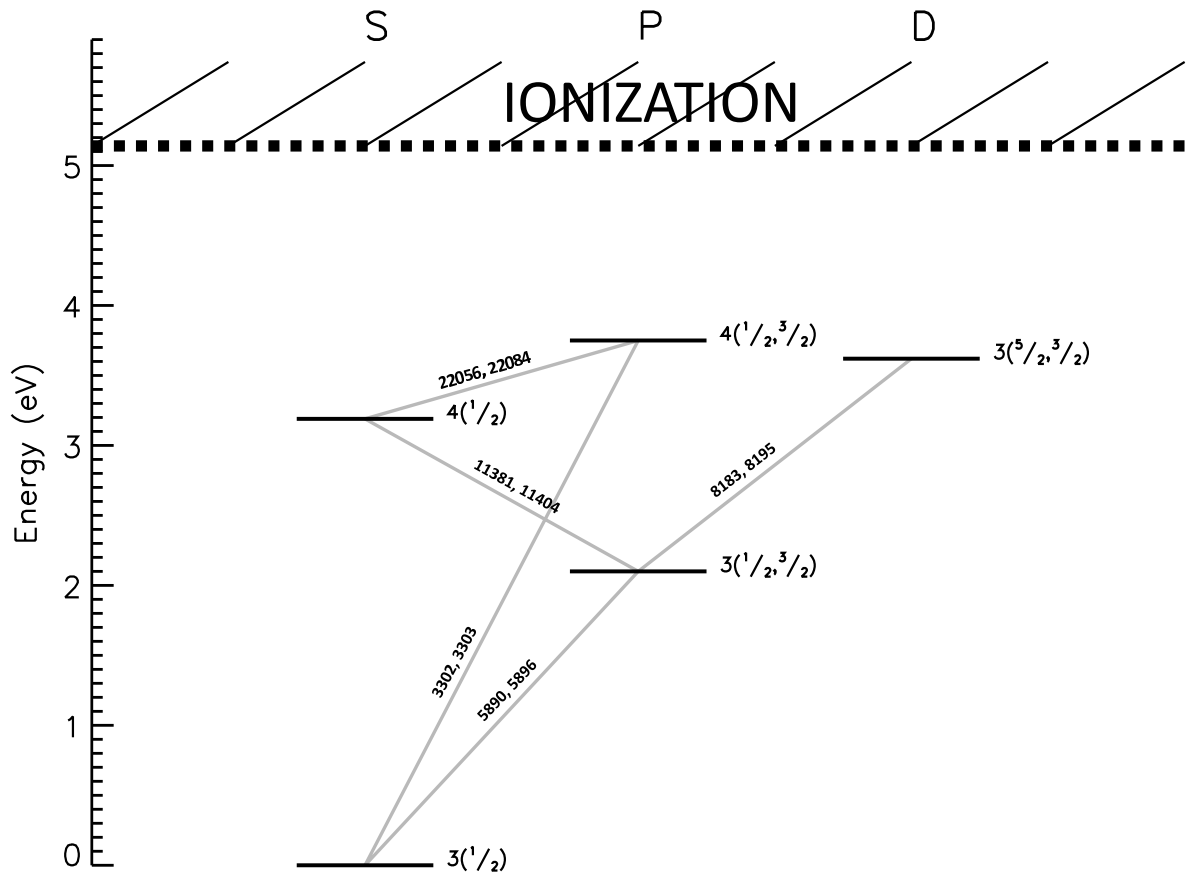}
\caption{Simplified \emph{NaI} atomic energy level diagram adapted from Bashkin \& Stoner (1975).  Transition wavelengths are in~\AA~.}
\label{elevels}
\end{figure}

\clearpage

\begin{figure}
\epsscale{1.0}
\plotone{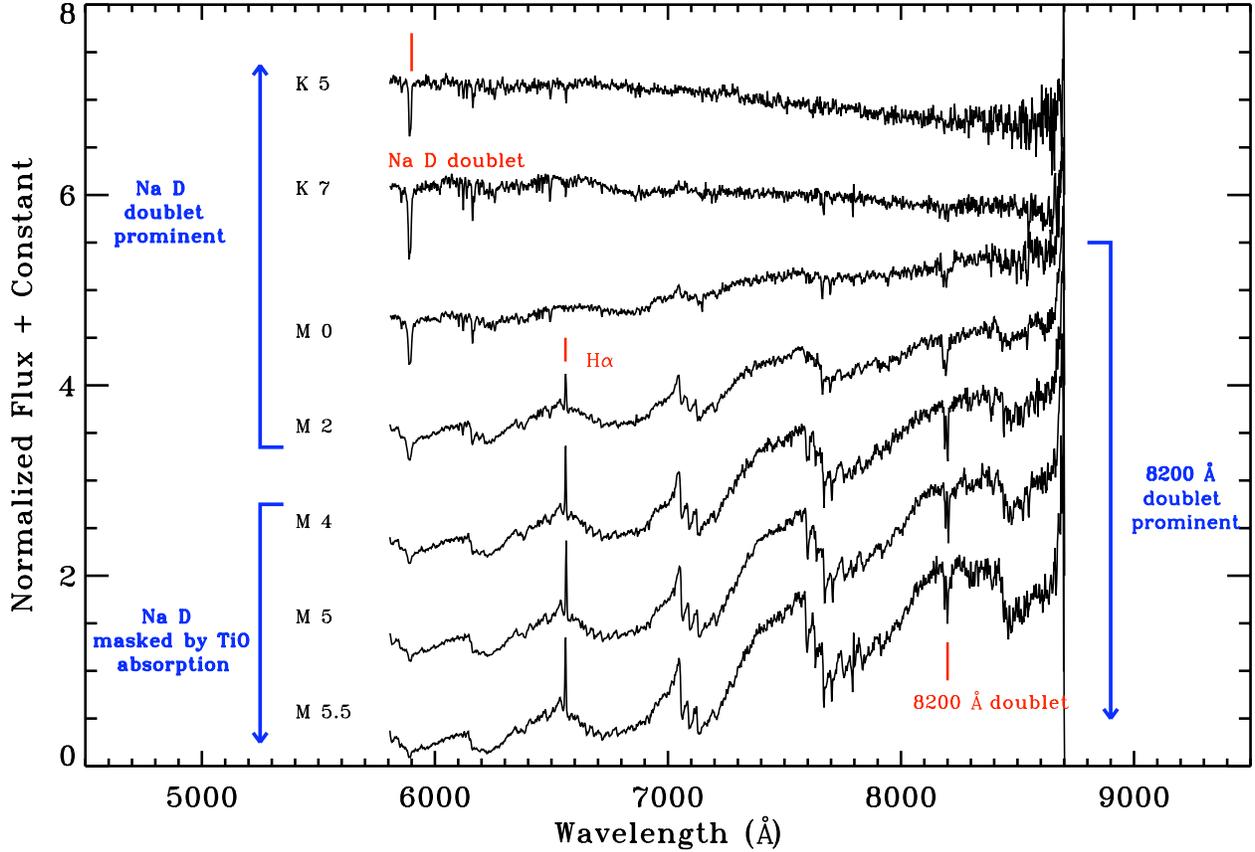}
\caption{3~\AA~resolution spectra of \emph{BPMG} candidate stars.  The spectra show that:  1) The unresolved \emph{Na D} doublet is detected in all spectra but is dominated by strong TiO molecular absorption for stars later than about M2.  \emph{Na D} \emph{EW} measurements would be unreliable for later M's.  2) The 8200~\AA~doublet is strong in M dwarfs; its \emph{EW} is measurable starting at \emph{SpTy} M0.  3) H$\alpha$ emission is common among mid-M dwarfs making it an unreliable youth indicator for those \emph{SpTy}'s.  A color version of this figure is available in the online journal.}
\label{example_spec}
\end{figure}

\clearpage


\begin{figure}
\epsscale{1.0}
\plotone{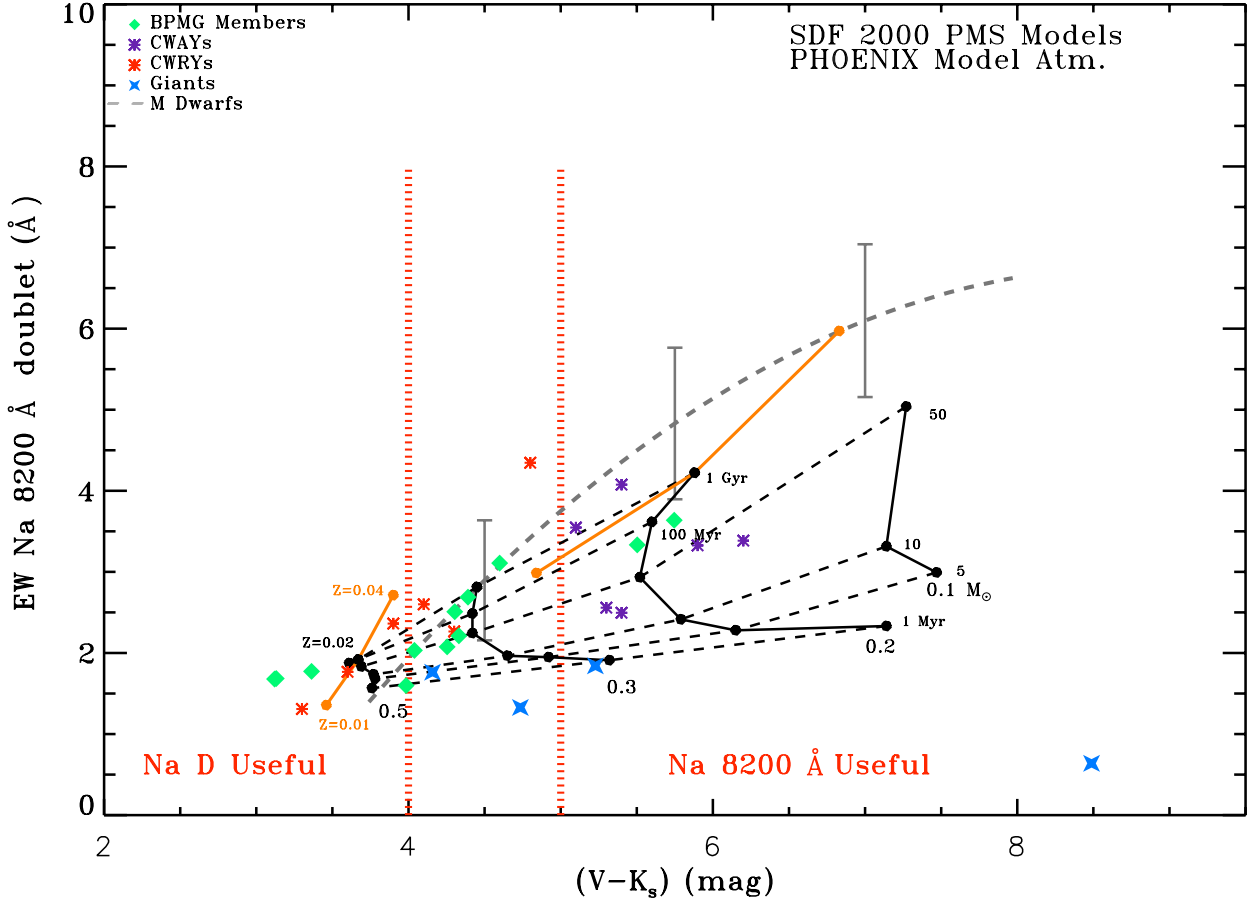}
\caption{Na 8200~\AA~\emph{EW} as a function of (\emph{V-K$_s$}) for the stellar samples described in the text (see key).  The dashed gray line is a second order polynomial fit to our \emph{HV} M dwarf sample (see \S 5 and Fig.~\ref{metal_plot}).  We used the SDF2000 \emph{PMS} evolution models to connect a stellar model of a specific mass and age with a model atmosphere of the corresponding (\emph{T$_{eff}$}, log($g$)) to plot isochrones and mass tracks.  The figure shows that: 1) (\emph{V-K$_s$}) $\gtrsim$ 5.0 \emph{BPMG} members and some candidates are clearly separated from the \emph{HV} dwarf sample and are consistent with young model isochrones, 2) Hence the 8200~\AA~doublet is a useful age diagnostic for stars of \emph{SpTy} $\sim$M4 and later (to the right of the second, thick, red, dotted line), where activity indicators become unreliable, and 3) for (\emph{V-K$_s$}) $<$ 5.0 the \emph{EW} is not useful because of quick convergence to the main sequence.  We also show the region where the \emph{Na D} line is useful as a luminosity class diagnostic (to the left of the first, red, dotted line).  The predicted change in \emph{EW} and color with metallicity is shown in orange.  A color version of this figure is available in the online journal.}
\label{82_SDF}
\end{figure}

\clearpage


\clearpage

\begin{figure}
\epsscale{1.1}
\plotone{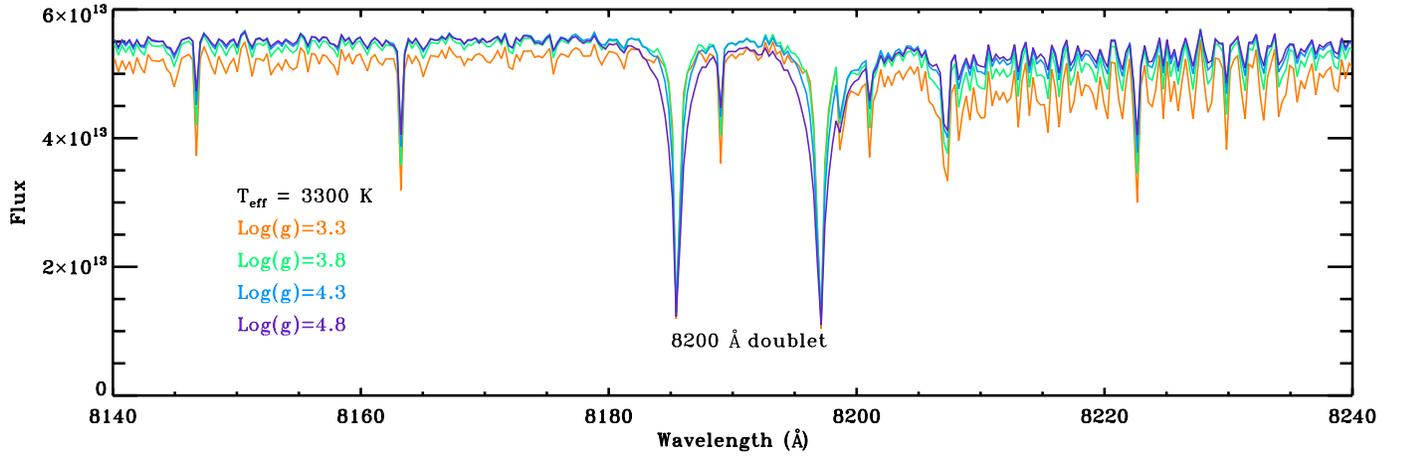}
\caption{Theoretical spectra in the region of the Na 8200~\AA~lines calculated using the PHOENIX stellar atmosphere code binned to a resolution of 3~\AA.  We present models for \emph{T$_{eff}$} = 3300 K over a range of log($g$).  The increase in line width with increasing log($g$) is evident.  A color version of this figure is available in the online journal.}
\label{syn_spec}
\end{figure}

\clearpage

\begin{figure}
\epsscale{1.0}
\plotone{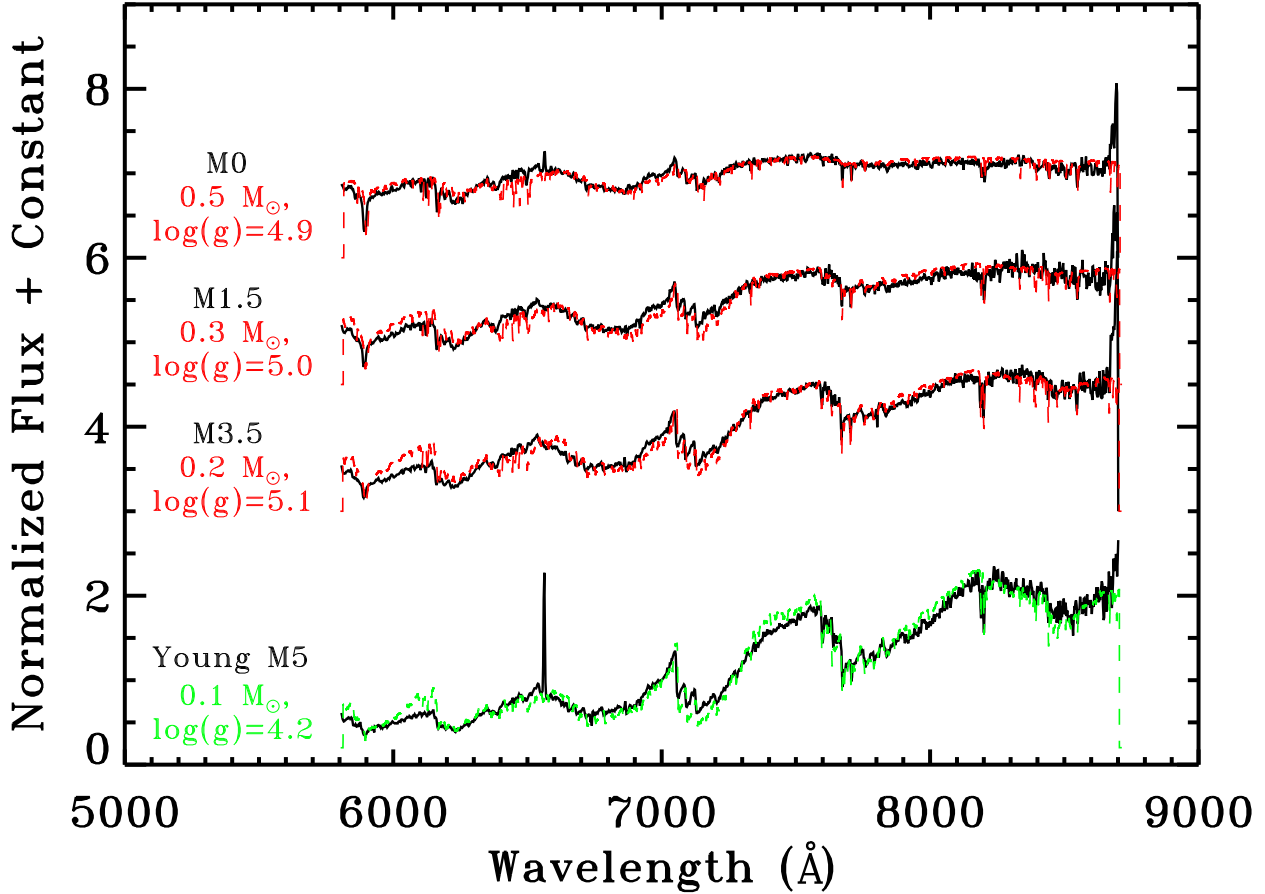}
\caption{Comparison of observed spectra to models across the observed wavelength range.  In the top of the figure observed (black) main sequence dwarfs are compared to 1 Gyr old PHOENIX model spectra with parameters derived from SDF2000 evolutionary models (red).  The models and observations generally match well but there are known discrepancies in some molecular bands in the models (TiO, $\sim$5900~\AA) due to incomplete line lists.  The bottom of the figure shows a probable young, mid-M dwarf compared to a 10 Myr old model spectrum (green).  The model is consistent with the observation and the agreement at the TiO band is improved; consistent with previous results (see R10).  Measured Na 8200~\AA~doublet \emph{EW} differences between observed spectra and their synthetic counterparts are less than the differences attributable to gravity at (\emph{V-K$_s$}) $\gtrsim$ 5.0.  The spike around 8700~\AA~in the observed spectra is attributable to instrumental effects.  A color version of this figure is available in the online journal.} 
\label{obs_comp}
\end{figure}

\clearpage

\begin{figure}
\epsscale{1.0}
\plotone{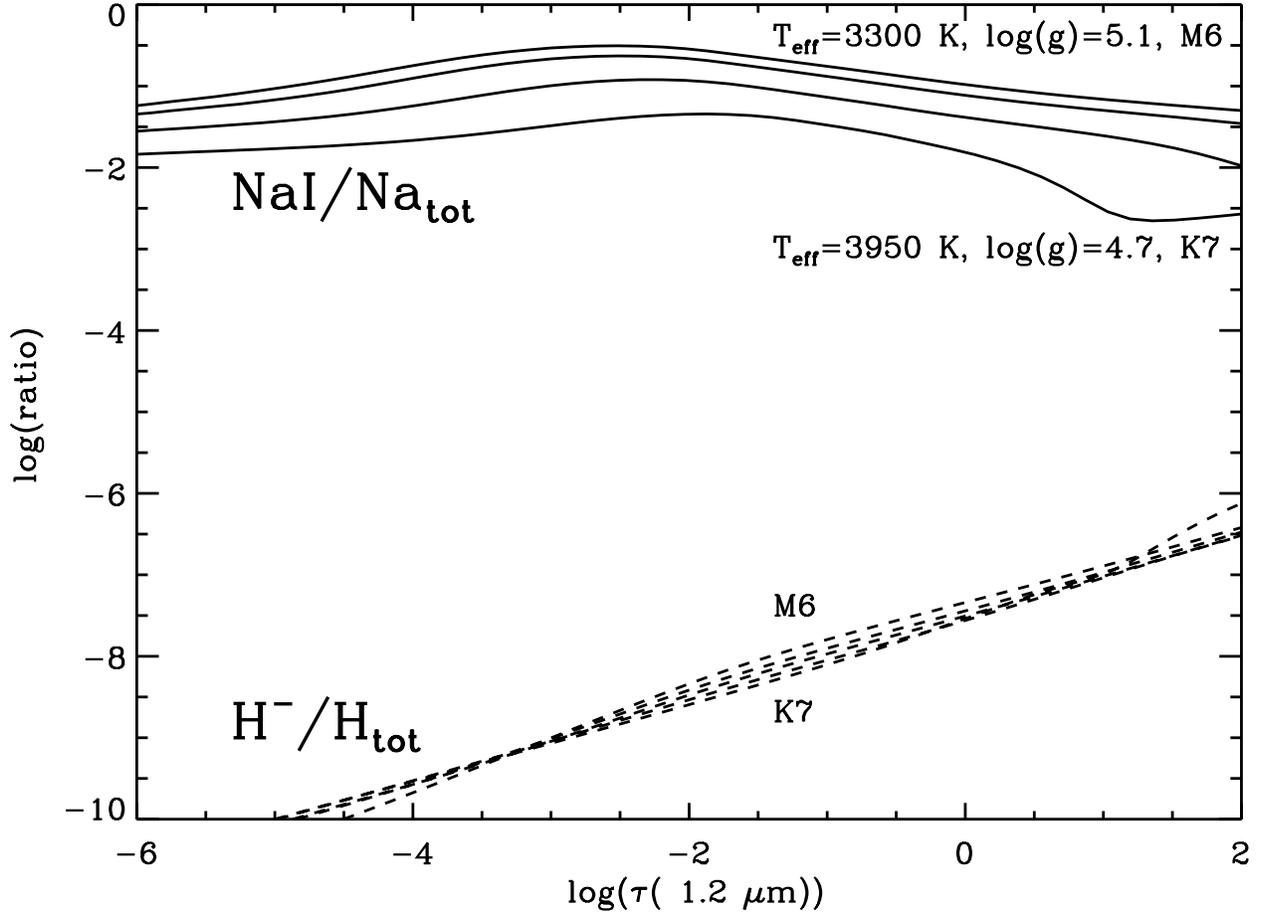}
\caption{Ratios of $NaI/Na_{tot}$ (solid) and $H^-/H_{tot}$ (dashed) as a function of $\tau$ for 1 Gy old stars.  The lines show the ratios for the K7- M6 \emph{SpTy} sequence (bottom line to top line) predicted from BCAH98 model data and PHOENIX model atmospheres.  In the top of the figure,  as \emph{T$_{eff}$} of the star decreases an increasingly larger fraction of Na remains neutral.  Thus the 8200 \AA~ doublet increases in strength in stars cooler than M0.  In the bottom of the figure, the H ratio remains roughly constant with \emph{T$_{eff}$} but increases substantially with increasing $\tau$.}
\label{abunrat}
\end{figure}

\clearpage

\begin{figure}
\epsscale{1.0}
\plotone{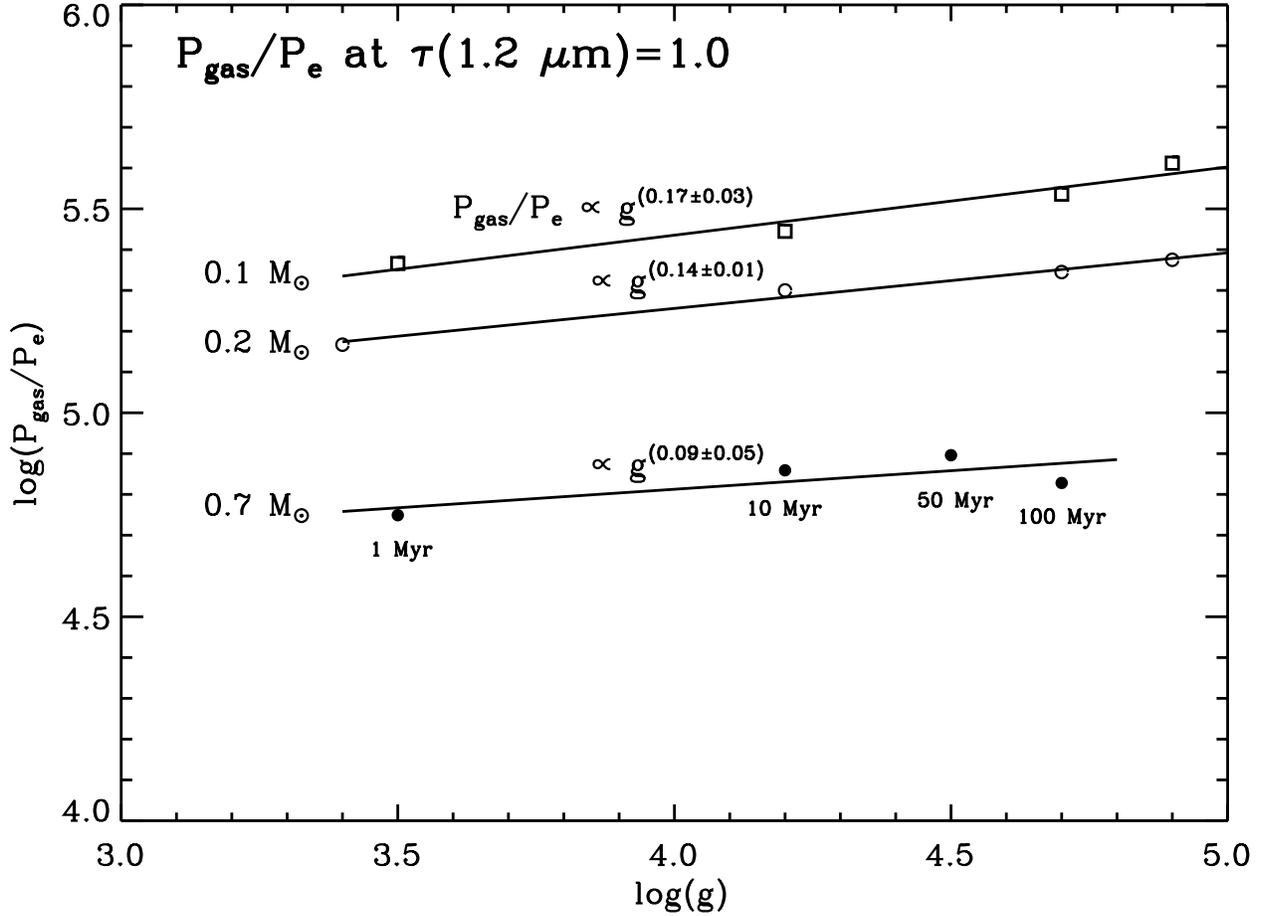}
\caption{$P_{gas}/P_{e}$ as a function of $g$ for our grid of masses and ages.  $P_{gas}/P_{e}$ increases with age and hence the Na doublets increase in strength.  The dependence of the pressure ratio on $g$ increases as stellar mass decreases.  This is consistent with our empirical result that the \emph{EW} of the 8200~\AA~doublet becomes a reliable youth indicator for dwarfs with \emph{SpTy} later than mid-M.}
\label{gasrat}
\end{figure}

\begin{figure}
\epsscale{1.0}
\plotone{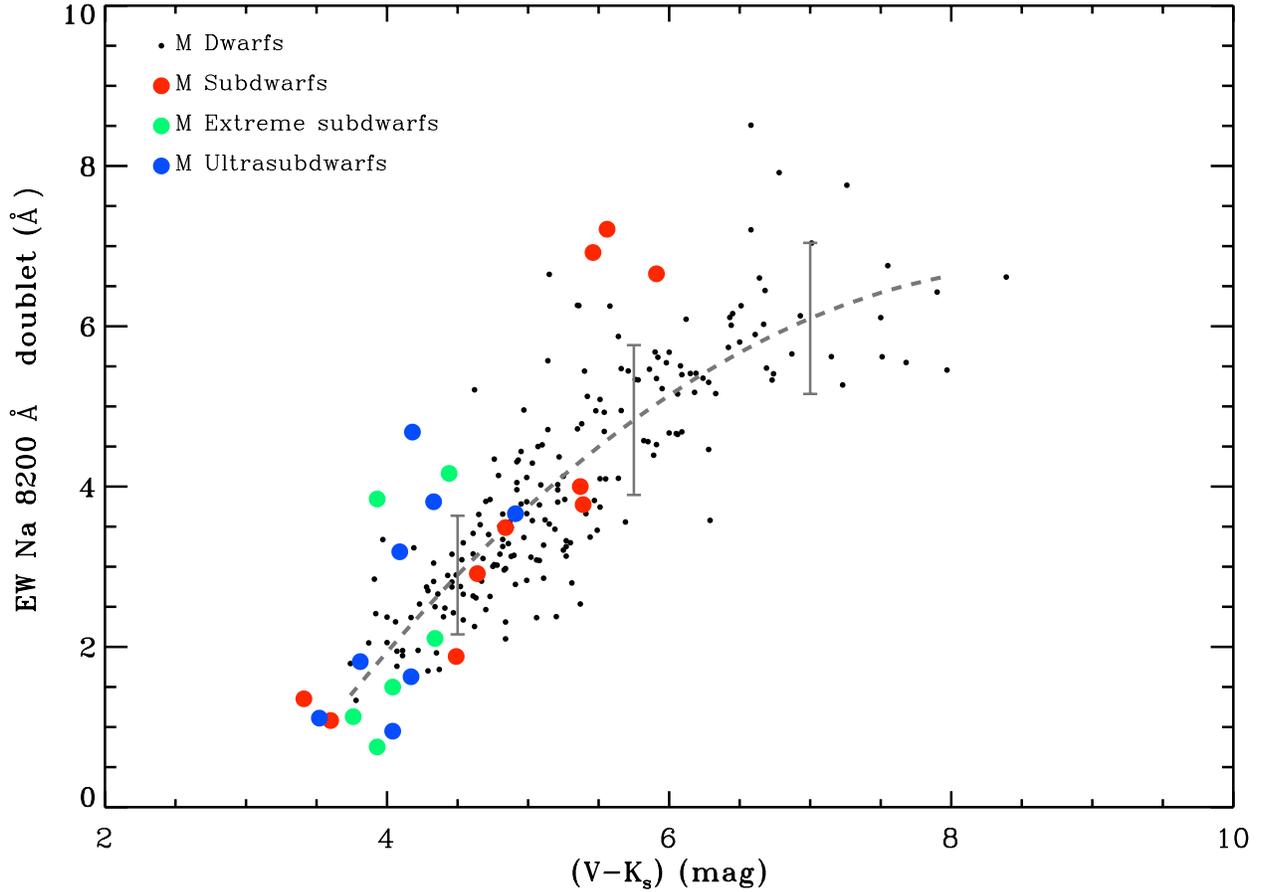}
\caption{\emph{EW} of Na 8200~\AA~doublet as a function of (\emph{V-K$_s$}) color for \emph{HV} M-dwarfs with samples of low-metallicity dwarfs from L07 (see key).  The gray, dashed line is a second order polynomial fit to the \emph{HV} M-dwarf data.  The uncertainty of the \emph{EW} measurements across the observed (\emph{V-K$_s$}) range were estimated by binning the measurements in increments of (\emph{V-K$_s$}) and taking the standard deviation in each bin.  The uncertainty increases toward later-type M dwarfs, probably a result of increased noise in fainter targets.   As metallicity decreases, dwarfs of like \emph{SpTy} are shifted toward smaller \emph{EWs} and bluer colors; consistent with predictions.  The analysis confirms that the Na 8200~\AA~doublet is useful as a youth diagnostic in low-mass M dwarfs for samples with homogeneous metallicity.  A color version of this figure is available in the online journal.}
\label{metal_plot}
\end{figure}

\end{document}